# BRIGHT 30 THz IMPULSIVE SOLAR BURSTS

P.Kaufmann[1,2], S.M. White[3], R. Marcon[4,5], A.S. Kudaka[1], D. P. Cabezas[1], M.M. Cassiano[1], C. Francile[6], L.O.T. Fernandes[1], R.F. Hidalgo Ramirez[1], M. Luoni[7], A. Marun[8], P. Pereyra[9], R. V. de Souza[1,2]

[1]*Center of Radio Astronomy and Astrophysics, Engineering School, Mackenzie Presbyterian University, São Paulo, SP, Brazil, e-mail: pierrekau@gmail.com*
[2]*Center of Semiconductor Components, State University of Campinas, Campinas, SP, Brazil.*
[3]*Air Force Research Laboratories, Space Vehicles Directorate, Albuquerque, NM, 87117, USA.*
[4]*"Gleb Wataghin" Physics Institute, State University of Campinas, Campinas, SP, Brazil.*
[5]*"Bernard Lyot" Solar Observatory, Campinas, SP. Brazil.*
[6]*Observatorio Astronómico Felix Aguilar, Universidad Nacional de San Juan, San Juan, Argentina.*
[7]*Instituto de Astronomia y Física del Espacio, CONICET-Universidad de Buenos Aires, Argentina.*
[8]*Instituto de Ciencias Astronomicas, de la Tierra y del Espacio, CONICET, San Juan, Argentina.*
[9]*Complejo Astronomico El Leoncito, CONICET, San Juan, Argentina.*

Corresponding author: Pierre Kaufmann, e-mail: pierrekau@gmail.com

## ABSTRACT

Impulsive 30 THz continuum bursts have been recently observed in solar flares, utilizing small telescopes with a unique and relatively simple optical setup concept. The most intense burst was observed together with a GOES X2 class event on October 27, 2014, also detected at two sub-THz frequencies, RHESSI X-rays and SDO/HMI and EUV. It exhibits strikingly good correlation in time and in space with white light flare emission. It is likely that this association may prove to be very common. All three 30 THz events recently observed exhibited intense fluxes in the range of $10^4$ solar flux units, considerably larger than those measured for the same events at microwave and sub-mm wavelengths. The 30 THz burst emission might be part of the same spectral burst component found at sub-THz frequencies. The 30 THz solar bursts open a promising new window for the study of flares at their origin.

## INTRODUCTION

Solar flares have been poorly explored in the terahertz to mid-IR range of frequencies. Observations at these wavelengths can help to resolve the source of the recently discovered mysterious burst component with flux spectra increasing through the sub-THz frequency range, and distinct from the well-known microwave flare emission that typically peaks at 10-20 GHz [for example *Kaufmann et al.* 2004, 2009*, Silva et al.* 2007*, Kaufmann,* 2012]. The THz flare emissions raise serious problems for interpretation. Early evidence on flare-associated hard X-rays known from a few rocket-based detectors measurements raised the first suggestions that associated radio emissions, known up to few tens of GHz, could be attributed to non-thermal synchrotron emission from high energy electrons, which led to the suggestion that the shorter-wavelength emission might be an extension of the synchrotron spectrum [*Stein and Ney*, 1963, *Shklovsky*, 1964]. This idea was critically evaluated by *Korchak* [1971] and *Brown* [1976] and did not survive, mostly because of the long burst time scales then believed to apply (10 to 1000 seconds) and because bremsstrahlung by lower energy electrons was sufficient to explain the observed X-rays [*Acton,* 1965]. The assumption of high energy non-thermal emission mechanisms at shorter wavelengths became again a likely possibility to explain bursts exhibiting subsecond time structures, with cm-mm fluxes increasing with frequency into the sub-THz regime, and time profiles well correlated to hard X-rays and gamma rays [*Kaufmann et al.,* 1985; *Kaufmann et al.,* 2004; *Kurt et al.,* 2010].

Several models have been proposed recently to explain the physical processes producing the sub-THz spectral burst component, most of them requiring the acceleration of particles to very high energies [for example, *Sakai et al.,* 2006, *Trottet et al.,* 2008, *Fleishman and Kontar,* 2010, *Krucker et al.,* 2013, *Klopf et al.,* 2014 and references therein]. More than one mechanism may be acting at the same time. While these processes are intended to explain the THz component, they do not account for the concurrent microwave component that is also observed.

The remarkable observational results obtained at sub-THz frequencies (0.2 and 0.4 THz) by the SST telescope at El Leoncito, Argentina Andes [*Kaufmann et al.,* 2008] have motivated the implementation of new telescopes to observe solar flares in the upper frequency band for which the terrestrial atmosphere becomes transparent, at 30 THz (the 8-14 µm band, also referred as mid-IR). The atmospheric transmission in this band is typically of the order of 95% at sea level, further improving with altitude, as it can be estimated from NASA ATRAN software simulations for atmosphere characterization with geographic locations and altitudes [Lord, 1992; ATRAN/SOFIA website, 2001]. It approaches to 100% at higher altitudes.

Past solar observations at 30 THz mostly refer to quiet Sun and quiescent regions [*Turon and Léna*, 1970, *Lindsey and Heasley,* 1981, *Gezari et al.,* 1999, Marcon *et al.,* 2008]. Possible flare activity at these frequencies was discussed with inconclusive observational results [*Hudson,* 1975], while *Xu et al.* [2004, 2006] detected white-light flares in the near-infrared at 1.56 microns. The detection of five minute oscillations has been suggested at mid-IR frequency bands [*Lindsey,* 1977]. In this paper we describe the instrumentation used for 30 THz observations, and after briefly reviewing two earlier events for comparison, we describe a new event observed on 2014 October 27 and discuss its implications.

## 30 THz SOLAR TELESCOPES

The first setup designed to detect solar flares at 30 THz was initially tested at "Bernard Lyot" Solar Observatory, Campinas, Brazil, using a Hale-type coelostat with two 20 cm flat mirrors reflecting radiation into a 15 cm diameter Newtonian telescope, feeding an early type Wuhan IR928 30 THz uncooled camera amorphous silicon 320x240 microbolometer array sensitive to about 0.5 K. It has been later installed at El Leoncito observatory, in Argentina Andes, utilizing a Jensch type coelostat with 30 cm flat mirrors, projecting radiation into a 15 cm diameter Newtonian telescope, for a number of observational campaigns [*Melo et al.,* 2006 and addendum in 2009, Marcon *et al.,* 2008, *Kaufmann et al.,* 2008, *Cassiano et al.,* 2010]. The setup has been later modified, forming a solar image of nearly 3/4 of the disk at the focal plane by a focal adjustment done by placement of a germanium lens in front of an uncooled vanadium oxide microbolometer array of a FLIR A20 8–15$\mu$m (30 THz) 320 × 240 pixel camera, sensitive to about 0.1 K [*Marcon et al.,* 2008, *Kaufmann et al.,* 2008]. The approximate band pass is basically set by the 8-14 µm atmospheric transmission window (i.e, 30 ± 8 THz) which is within the well known germanium lens transmission band from 2-14 µm. To ensure conclusively that near-IR in the several wavelength window bands shorter than 5.2 µm and visible radiation are effectively suppressed, tests were performed by interposing a specially designed >3$\mu$m low-pass filter made by Tydex LCC company (Saint Petersburg, Russia) when pointing to the Sun (see Figure 1).

Frames are obtained with a cadence of 5 s$^{-1}$. Recently a similar 30 THz solar flare telescope was installed at the top of a building at Mackenzie Presbyterian University, in São Paulo, Brazil [*Kudaka et al.,* submitted to Solar Phys, 2015]. It uses a Hale-type coelostat with two 20 cm flat mirrors feeding radiation into a similar 15 cm diameter Newtonian telescope, also coupled to the Wuhan IR928 30 THz camera. Observations are again carried out with a cadence of 5 frames s$^{-1}$. Both 30 THz telescopes have a spatial resolution of about 15 arcseconds, corresponding to the Airy disk set by the primary reflector's diffraction angle limit, sometime called "photometric" beam. It corresponds to about four 4" pixels on the detector and about 16 pixels over the Airy disk area.

The observed 30 THz burst excess brightness temperature, compared to the photosphere level, was calibrated from the brightness temperature difference between the effective temperature of the photosphere and the sky equal to about 4700 K, assuming the disk temperature at 30 THz to be about 5000 K [*Turon and Léna,* 1970; *Gezari et al.,* 1999] and the sky to be at the ambient temperature of about 300 K (due to the emission from mirrors, germanium lens and blocking devices interposed) [*Kaufmann et al.,* 2013]. The burst excess brightness temperature is determined for a position close to the bright burst centroid of emission, on images averaged over one second (i.e., 5 frames) subtracted from the level measured at neighboring quiet photospheric regions, or at the same burst position before the event. The temperature scale has been verified to be approximately correct, since it provides a sunspot umbra temperature decrement within the range measured by other authors (i.e., -500 K to -1000 K) [*Turon and Léna,* 1970, *Gezari et al.,* 1999, *Marcon et al.,* 2008].

For sources that are no bigger than the "photometric" beam (i.e. about 15 arcseconds), we can estimate the 30 THz flux density using the well known relationship $\Delta S = 2$ k $\Delta T/A_e$ [*Kraus,* 1986], where $\Delta S$ is the flux density, k the Boltzmann constant, $\Delta T$ the excess brightness temperature due to the burst, and $A_e$ the aperture effective area. This equation is derived assuming the Rayleigh-Jeans (RJ) approximation to Planck's law for the 30 THz emission of the solar disk. Indeed for the solar disk brightness temperature T ≈ 5000 K, the exponent in Planck's law is hf/kT<1, where h is the Planck constant, f = 3 10$^{13}$ Hz the frequency. Compared to the complete Planck's equation the RJ brightness becomes over-estimated by about 15% for T = 5000 K and less when excess temperature is added during bursts. The estimated effective area $A_e$ is about 0.0028 m$^2$, similar for both telescopes, after discounting all the losses [*Kaufmann et al.,* 2013; *Kudaka et al.,*submitted to Solar Physics, 2015; *Miteva et al.*, submitted to Astron. Astrophys., 2015].

**THE FIRST OBSERVATIONS OF IMPULSIVE 30 THz SOLAR BURTS**

The first impulsive 30 THz solar burst was observed on March 13, 2012 (SOL2012-03-13T17:24) at El Leoncito and simultaneously in a wide radio range, sub-THz, SDO EUV and HMI white light, and FERMI hard X-rays [*Kaufmann et al.*, 2013]. It was associated with a GOES class M8 flare occurring in NOAA active region AR1429 at heliographic coordinates N18W50. The 30 THz time profile was found to be consistent with the impulsive profiles obtained at microwaves by the US AFRL RSTN network, with the 50-100 and 300-540 keV hard X-ray ranges from FERMI, the El Leoncito 45 and 90 GHz solar polarimeters, and SST at 212 GHz. At 405 GHz the SST flux was below the detection limit. There was a strikingly good time and space coincidence between the 30 THz brightening and the white-light flare (WLF) derived from SDO/HMI images. The 30 THz peak flux was of about 12000 SFU (with an uncertainty of order 25%, i.e. ± 3000 SFU; 1 SFU =10$^{-22}$ Wm$^{-2}$Hz$^{-1}$), which was almost three decades more intense than the fluxes measured at microwave and sub-THz frequencies for that event [see *Kaufmann et al.,* 2013].

Another intense impulsive 30 THz solar burst was observed by the São Paulo telescope on August 1$^{st}$, 2014 (SOL2014-08-01T14:47), during a GOES class M2 soft X-ray on NOAA AR 2130 (S09E35). Detailed analysis of this event has been given by *Miteva et al.*[submitted to Astron. Astrophys., 2015]. It was simultaneously observed in a wide range of radio wavelengths, from metric to microwaves (by Orfees/Nançay, France; Ondrejov, Czech Republic and US AFRL RSTN network), sub-THz by SST, Hα by HASTA (in Argentina) and at EUV wavelengths by SDO. The 30 THz time profile was similar to the March 13, 2012 burst. However no obvious white light emission could be seen in the SDO/HMI images of this event. The 30 THz flare source position coincides with brightenings observed in the Hα, the SDO 304 and 1700 Å flare images. The 30 THz excess brightness temperature at the peak was of about 4 % above the photosphere corresponding to 20000 SFU using the same analysis procedure described above, with an uncertainty of ± 25%. The emission spectrum has the same trend as found for the March 13, 2012 burst [*Kaufmann et al.*, 2013], exhibiting two continuum spectral components, one with a peak emission at microwaves and another one in the THz range, with the 30 THz flux almost two orders of magnitude larger than microwaves, and three orders of magnitude larger than the upper limit of the sub-THz emissions [see *Miteva et al.,* submitted to Astron. Astrophys., 2015].

**4. 30 THZ IMPULSIVE BURST WITH THE X2 OCTOBER 27, 2014 FLARE**

A GOES class X2.0 soft X-ray burst occurred on October 27, 2014 (SOL2014-10-27T14:22) in NOAA AR 2192 (S18W57). It was well observed at 30 THz by the São Paulo telescope during several minutes around the maximum of the event. Moving clouds in the sky prevented the recording of a complete time profile. The event was also observed by HASTA (Argentina) at Hα, by SDO in EUV and HMI white light, at sub-THz by SST (El Leoncito) and in microwaves by US AFRL RSTN. A white light flare enhancement at two distinct sites has been observed. The SDO/HMI, visible, Hα and 30 THz images are shown in Figure 2. The 30 THz heliographic position is obtained by aligning to the sunspots on the disc, with an accuracy of 10-15 arcseconds. The upper panels exhibit two white light spots from SDO/HMI with positions indicated in the visible disk, the lower panels show the Hα and the 30 THz brightening at about the time of maximum burst emission. The 30 THz flux has been derived from the same method described in Section 2, being about 35000 SFU (± 25%) (corresponding to a brightness temperature enhancement of about 7% above the photospheric level, or 330 K).

This event was observed at higher microwave frequencies by RSTN, RHESSI and at two sub-THz frequencies by SST. The SST time profiles show a clear spike at 0.4 THz and 0.2 THz approximately coincident in time with WLF peaks of the north and south sources, derived from SDO/HMI shown in Figure 3. The SDO/HMI time profiles have been obtained following the same standard procedures used before [*Kaufmann et al.,* 2013]. HMI white-light images are derived by fitting the continuum intensity levels on either side of the Fe I absorption line at 6173.34 Å used by HMI for helioseismology and magnetic field measurements. For the analysis of white-light flare emission we use images from which a pre-flare image has been subtracted. The light curves for sources S1 and S2 were obtained by averaging on angular scales of 3" x 3". The SST plots are in units of antenna temperature corrected for atmospheric transmission for beams 4 (0.2 THz) and 5 (0.4 THz). The antenna temperature corresponds approximately to the solar brightness temperature multiplied by the respective beam efficiency [*Kraus*, 1986], above which the burst excess temperature is added. SST operates using the multiple partially overlapping beams technique [*Georges et. al.,* 1989; *Giménez de Castro et al.,* 1999; *Kaufmann et al.,* 2008] to further correct the antenna temperatures for the burst source relative position with respect to the beams (i.e., a cluster of three partially overlapping 4 arc-minute beams at 0.2 THz and one 2 arc-minute beam at 0.4 THz at the center, plus overlapping 0.2 and 0.4 THz beams displaced from the cluster by 8 arc-minutes for comparison). However the correction algorithm becomes inaccurate for weak burst, when beams are near the solar limb. The spike was observed at about 14:22 UT only in beams 4 (0.2 THz) and 5 (0.4 THz) exhibiting antenna temperatures corrected for atmosphere transmission of about 25 and 220 K. There was no significant signal at beams 2 and 3, which were pointing partially on the solar limb, and no signal on the two comparison beams that were pointing away from the limb. Therefore the calculated flux densities of about 10 SFU and 150 SFU at 0.2 and 0.4 THz, respectively, should be considered to be lower limit values, well within uncertainties of the order of ±25%, due to other corrections.

RHESSI time profiles at low to high energy ranges, plotted at the top of Figure 3, exhibit a time coincident impulsive enhancement at the energy range 50-100 keV, and to a lesser contrast, also at 25-50 keV.

The maximum emission time of 14:22 UT coincides with RSTN reports of the peak times at microwave frequencies. The complete burst spectrum at maximum is shown in Figure 4. The spectral trends are similar to the events mentioned in Section 3 (i.e., SOL2012-03-13T17:24 and SOL2014-08-01T14:47) [*Kaufmann et al.,* 2013; *Miteva et al.,* submitted to Astron. Astrophys., 2015], exhibiting one component with a maximum at microwaves, and another component in the THz range. The 30 THz peak flux density of 35000 sfu estimated for this burst, however, is qualitatively larger than the two previous bursts (i.e., 12,000 for the March 13, 2012 burst and 20,000 for the August 1[st], 2014 burst, respectively – see Section 3 above).

**DISCUSSION**

The 30 THz impulsive solar bursts observed in connection with M flares and with an X flare discussed here suggest that the sub-THz spectral component with fluxes increasing with frequency, distinct from microwaves, may be extended to 30 THz, exhibiting much stronger fluxes of order $10^4$ SFU. Of course, the full demonstration of this suggestion requires observations at more THz frequencies, which are not available. The whole radio-microwave to THz observed spectral trends appear very similar for the three bursts. For two events there were white-light flares coincident in time and space. These were well observed impulsive events, all three bursts exhibited fluxes nearly one-two orders of magnitude larger than at microwaves and at sub-THz frequencies.

The THz burst spectral component is clearly distinct from the typical non-thermal radio burst with a peak at about 9 GHz, which is usually attributed to gyrosynchrotron emission from mildly relativistic electrons [*Dulk,* 1985; *Bastian et al.,*1998]. If we assume that the 30 THz and the sub-THz observed emissions originate from the same source, the positive spectral index close to 2 is qualitatively consistent with optically thick emission that could be either thermal or non-thermal. However the close time coincidence with the sub-THz impulsive burst, with RHESSI harder X-rays only (50-100 keV), EUV bright flare, and the white light two-source brightening may favor a non-thermal origin for the emission process. Such a mechanism was inferred from white light flares observed in the near-infrared range of wavelengths [*Xu et al.,* 2006]. They found that the WLF time synchronism improves for harder hard X-rays strongly suggesting the acceleration of high energy particles as the energy source [originally suggested by *Najita & Orrall, 1970*]. Similarly, *Martínez Oliveros et al.* [2012] found that the heights of WL emission and 30-80 keV hard X-rays from a limb event were very similar, implying that nonthermal electrons can penetrate to the layer from which WL emission is seen. In the case of the 2012 March 13 flare [*Kaufmann et al., 2013*], differences in the behavior of the 30 THz and white-light emission were noted. An alternative mechanism of radiative back-warming has been proposed, in which higher layers are heated by non-thermal electrons and they then heat the deeper temperature-minimum layers via Balmer and Paschen continua [*e.g., Machado, Emslie and Avrett,* 1989*; Ding et a.,* 2003*; Xu et al.,* 2004]. This mechanism has several implications that should be tested, and in particular the fact that the white light emission is a secondary product would seem to exacerbate the difficulty explaining the energy radiated in white light events [e.g., *Neidig,* 1989].

With a sample of just 3 events so far, the absence of white-light emission from one of the flares may or may not be significant, but the fact that the 30 THz emission in the 2012 March 13 flare was more readily detectable than the white light emission seen in HMI [*Kaufmann et al. 2013*] suggests that the near-infrared may be the best wavelength region, and certainly more sensitive than white-light continuum, to study emission from the low chromosphere/temperature minimum region.

The 30 THz fluxes in these events are 1-2 orders of magnitude smaller than fluxes roughly estimated for white light flares usually attributed to very large flares, i.e. $10^5$-$10^6$ SFU [*Stein and Ney,* 1963; *Ohki and Hudson*, 1975; *Neidig and Cliver,* 1983; *Cliver and Svalgaard,* 2004]. The WLF flux estimates have large uncertainties. They usually refer to flaring areas with scale size of 10 arcseconds in average [*Neidig and Cliver,* 1983]. Similarly the flux densities are derived from the 30 THz brightness temperature increments relative to the photosphere diluted over relatively large photometric beam (i.e., the 15 arcsecond diameter Airy disk). We note that *Couvidat et al.* [2012] discussed issues with HMI white-light data due to the fact that the white-light image is generated from 12 filtergrams (2 polarizations at each of 6 wavelengths) of the Fe I 6173.34 Å line taken at different times over the 45s sampling period. Analysis assumes a Gaussian profile for the line and derives the continuum level (in the vicinity of the Fe line) as well as the Doppler shift, line width and other properties as linear sums over the 6 wavelengths observed in each pixel. Time variability within the 45 s measurement interval, which we would certainly expect in the rise phase of an impulsive flare, can distort the results, and the assumption of a Gaussian line profile may not be valid during a flare. In the case discussed by *Martinez Oliveros et al.* [2014], HMI data showed spurious blue-shifts of the Fe line. In addition, temporal interpolation involving negative coefficients in adjacent 45 s intervals can result in spurious "black-light" flares [*Martinez Oliveros et al.* 2011]. These factors complicate the interpretation of the HMI white-light continuum images, and presumably introduce a large uncertainty in the corresponding white-light fluxes.

**CONCLUDING REMARKS**

The first solar observations of impulsive 30 THz bursts exhibit intense fluxes, in the range of $10^4$ SFU, nearly one to two orders of magnitude larger than corresponding emissions at microwaves and sub-THz frequencies. A "double spectral" feature becomes evident in the spectrum, with one component showing typical radio fluxes peaking at microwave frequencies, and the other component corresponding to the recently identified sub-THz to THz component whose flux increases with frequency. "Double spectral" features in the cm-mm range of frequencies have been known since the early seventies [as for example *Croom*, 1971; *Akabane etl al.,* 1973; *Shimabukuro*, 1973; *White et al*., 1992 and several others, reviewed by *Kaufmann,* 2012]. This spectral feature might be similar to the sub-THz to THz component, with the transition frequency minimum shifted towards higher frequencies. To better understand the nature of these spectral index inflections, observations at more frequencies are needed, especially in the 0.4 to 30 THz range of frequencies.

White light flares were observed for two events, coincident in time and space. The 30 THz fluxes are of about one to two orders of magnitude smaller than estimates of the white light flare flux for very large events. The analysis of the events here strongly suggests that sub-THz, 30 THz and white-light flare emissions may arise from the same bursting source, having close physical connections to be further studied. More studies are required to conclude whether these results favor the non-thermal heating origin for the 30 THz emission, as also suggested by the spatial correlation of WLF and near-infrared emission with hard X-rays [*Najita & Orrall, 1970; Xu et al.,* 2006; *Martínez Oliveros et al*., 2012], or thermal, backwarming processes [*e.g., Machado, Emslie and Avrett,* 1989*; Ding et al,* 2003*; Xu et al.,* 2004], or some other mechanism.

The frequency of maximum THz emission remains unknown; it can be at frequencies lower or higher than 30 THz, bringing constraints on the energy of particles and magnetic field involved. New experiments are currently being implemented to complete the spectral description at intermediate THz frequencies, from space (3 and 7 THz) and from the ground (0.85 and 1.4 THz) [*Kaufmann et al*., 2014*, Kaufmann,* 2015]. The observations and studies of solar bursts in the THz range open a new area of research connected to energetic processes at the origin of solar flares. The El Leoncito and São Paulo 30 THz telescopes are undergoing up-grading by replacing the cameras with more sensitive models that are expected to bring improvements in sensitivity by a factor of 5 for both instruments. They will be operated regularly, during nearly 300 clear days per year at El Leoncito, and 120 clear days per year at São Paulo.

*Acknowledgements*

The authors are grateful for the helpful remarks given by two anonymous reviewers that have improved considerably the article presentation. This research was partially supported by Brazilian agencies FAPESP (Proc. 2013/24177-3), CNPq, and Mackpesquisa; US AFOSR and Argentina CONICET. There are no restriction to use or access data presented here. Potential users may access data presented here corresponding directly with the principal author.

## References

Acton, L.W. (1965), Contribution of Characteristic X-rays to the Radiation of Solar Flares. *Nature, 207*, 737-738.

Akabane, K., Nakajima, H., Ohki, K., Moriyama, F., Miyaji, T. (1973). A Flare-Associated Thermal Burst in the mm-Wave Region. *Solar Phys., 33*, 431-437

Atran/SOFIA website (2001). https://atran.sofia.usra.edu/cgi-bin/atran/atran.cgi

Bastian, T. S., Benz, A. O. and Gary, D. E. (1998). Radio Emission from Solar Flares. *Ann. Rev. Astron. Astrophys., 36*, 131-188.

Brown, J.C. (1976). The Interpretation of Hard and Soft X-rays from Solar Flares. *Phil. Trans. Roy. Soc. London, 281*, 473-490.

Cassiano, M. M., Kaufmann, P., Marcon, R., Kudaka, A. S., Marun, A., Godoy, R., Pereyra, P., Melo, A. and Levato, H. (2010). Fast Mid-IR Flashes Detected During Small Solar X-Ray Bursts. *Solar Phys., 264,* 71-79.

Cliver, E. W. and Svalgaard, L. (2004). The 1859 Solar-Terrestrial Disturbance And the Current Limits of Extreme Space Weather Activity. *Solar Phys., 224,* 407-422.

Couvidat, S., Rajaguru, S. P., Wachter, R., Sankarasubramanian, K. Schou, J. and Scherrer, P. H. (2012). Line-of-Sight Observables Algorithms for the Helioseismic and Magnetic Imager (HMI) Instrument Tested with Interferometric Bidimensional Spectrometer (IBIS) Observations. *Solar Phys., 278,* 217-240.

Croom, D.L. (1971). Solar Microwave Bursts as Indicators of the Occurrence of Solar Proton Emission. *Solar Phys., 19*, 150-170.

Ding, M. D., Liu, Y., Yeh, C.-T., and Li, J. P. (2003). Interpretation of the infrared continuum in a solar white-light flare. *Astron. Astrophys., 403*, 1151-1156.

Dulk, G. A. (1985). Radio emission from the sun and stars. *Ann. Rev. Astron. Astrophys. 23,* 169-224.

Fleishman, G. D. and Kontar, E. P. (2010). Sub-Thz Radiation Mechanisms in Solar Flares. *Astrophys; J. Letters, 709,* L127-L132.

Georges, C.B., Schaal, R.E., Costa, J.E.R., Kaufmann, P. and Magun, A. (1989). 50 GHz Multi-Beam Receiver for Radio Astronomy. *Proc. 2nd SBMO–International Microwave Symposium*, São Paulo, Brazil, IMOC/IEEE–MTT Cat. N89th0260-0, V. II, 447.

Gezari, D., Livingston, W. and Varosi, F. (1999). Thermal Structure in Sunspots and Dynamic Solar Infrared Granulation Imaged at 4. 8, 12. 4, and 18. 1 Microns. High

Resolution Solar Physics: Theory, Observations, and Techniques, *ASP Conference Series #183*. Eds. T. R. Rimmele, K. S. Balasubramaniam, and R. R. Radick, p.559.

Giménez de Castro, C.G., Raulin, J.-P., Makhmutov, V.S., Kaufmann, P. and Costa, J.E.R., (1999). Instantaneous positions of microwave solar bursts: Properties and validity of the multiple beam observations, *Astron. Astrophys.*, *140*, 373-382.

Hudson, H. S. (1975). The solar-flare infrared continuum - Observational techniques and upper limits. *Solar Phys., 45,* 69-78.

Kaufmann, P., Correia, E., Costa, J. E. R., Vaz, A. M. Z. and Dennis, B. R. Solar burst with millimetre-wave emission at high frequency only. (1985). *Nature, 313,* 380-382.

Kaufmann, P., R. J., De Castro, C. G., Levato, H., Gary, D. E., Costa, J. E., Marun, A., Pereyra, P., Valio, A. and Correia, E. (2004). A new solar burst spectral component emitting only in the terahertz range. *Astrophys.J. 603,* L121-L124.

Kaufmann, P., Levato, H., Cassiano, M. M., Correia, E., Costa, J. E., Giménez de Castro, C. G., Godoy, R., Kingsley, R. K., Kingsley, J. S., Kudaka, A. S., Marcon, R., Martin, R. and Marun, A. (2008). New telescopes for ground-based solar observations at sub-millimeter and mid-infrared. *Proc. of SPIE,* 70120L, 8pp

Kaufmann, P., Giménez de Castro, C. G., Correia, E., Costa, J. E., Raulin, J. P. and Válio, A. S. (2009). Rapid pulsations in sub-THz solar bursts. *The Astrophysical Journal, 697,* 420-427.

Kaufmann. P. (2012), Observations of Solar Flares from GHz to THz Frequencies, *Astrophys. Space Sci. Proc., 30,* 61, Springer-Verlag Berlin Heidelberg.

Kaufmann, P., White, S. M., Freeland, S. L., Marcon, R., Fernandes, L. O., Kudaka, A. S., de Souza, R. V., Aballay, J. L., Fernandez, G., Godoy, R., Marun, A., Valio, A., Raulin, J.-P. and Giménez de Castro, C. G. (2013). A Bright Impulsive Solar Burst Detected at 30 THz. *Astrophys. J, 768,* 134-143

Kaufmann, P., Marcon, R., Abrantes, A., Bortolucci, E. C., Fernandes, L. O., Kropotov, G. I., Kudaka, A. S., Machado, N., Marun, A., Nikolaev, V., Silva, A., da Silva, C. S. and Timofeevsky, A. (2014). THz photometers for solar flare observations from space. *Exp. Astron., 37,* 879-598.

Kaufmann, P. (2015). Space and ground-based new tools for Thz solar flare observations. *26th International Symposium on Space Terahertz Technology,* Cambridge, MA, 16-18 March.

Klopf, J. M., Kaufmann, P., Raulin, J.-P. and Szpigel, S. (2014). The Contribution of Microbunching Instability to Solar Flare Emission in the GHz to THz Range of Frequencies. *Astrophys. J., 791,* 31, 11pp.

Korchak, A.A. (1971). On the Origin of Solar Flare X-Rays. *Solar Phys., 18,* 280-304.

Kraus, J. D. (1986). *Radio Astronomy* (2nd ed.; Powell, OH: Cygnus-Quasar Books), Section 6–2.

Krucker, S., Giménez de Castro, C. G., Hudson, H. S., Trottet, G., Bastian, T. S., Hales, A. S., Kasparová, J., Klein, K.-L., Kretzschmar, M., Lüthi, T., Mackinnon, A., Pohjolainen, S. and White, S. M. (2013). Solar flares at submillimeter wavelengths. *Astronon Astrophys. Rev., 21,* article id.58.

Kurt, V. G., Yushkov, B.Yu., Kudela, K. And Galkin, V.I. (2010). High-energy gamma radiation of solar flares as an indicator of acceleration of energetic protons *Cosmic Research, 48,* 70-79.

Lindsey, C. A. (1977). Infrared continuum observations of five-minute oscillations. *Solar Phys., 52,* 263-281.

Lindsey, C. and Heasley, J. N. (1981). Far-infrared continuum observations of solar faculae. *Astrophys. J., 247,* 348-353.

Lord, S.D. (1992). A new software tool for computing Earth atmosphere transmission of near- and far- infrared radiation, NASA Technical Memorandum no. 103957.

Machado, M.E., Emslie, A.G., Avrett, E.H. (1989). Radiative backwarming in white-light flares. *Solar Phys., 124,* 303-317.

Marcon, R., Kaufmann, P., Melo, A. M., Kudaka, A. S. and Tandberg-Hanssen, E. (2008). Association of Mid-Infrared Solar Plages with Calcium K Line Emissions and Magnetic Structures. *Publ.Astron. Soc. Pacific, 120,* 16-19.

Martínez Oliveros, J.-C., Couvidat, S., Schou, J., Krucker, S., Lindsey, C., Hudson, H.S., and Scherrer, P. (2011). Imaging Spectroscopy of a White-Light Solar Flare. *Solar Phys, 269,* 269-281.

Martínez Oliveros, J.-C., Hudson, H.S., Hurford, G.J., Krucker, S., Lin, R.P., Lindsey, C., Couvidat, S., Schou, J. and Thompson, W.T. (2012). The Height of a White-Light Flare and its Hard X-ray Sources. *Astrophys. J., 753,* L26 (5pp).

Martínez Oliveros, J.-C., Lindsey, C., Hudson, H.S., and Buitrago Casas, J. C. (2014). Transient Artifacts in a Flare Observed by the HMI on SDO. *Solar Phys., 289,* 809-819.

Melo, A., Kaufmann, P., Kudaka, A. S., Raulin, J.-P., Marcon, R., Marun, A., Pereyra, P. and Levato, H. (2006). A New Setup for Ground-based Measurements of Solar Activity at 10 mum. *Publ.Astron. Soc. Pacific, 118,* 1558-1563.

Melo, A., Marcon, R., Kaufmann, P., Kudaka, A. S., Marun, A., Pereyra, P., Raulin, J-P. and Levato, H. (2009). Addendum: "A New Setup for Ground-based Measurements of Solar Activity at 10 mum" (PASP, 118, 1558 [2006]). *The Publications of the Astronomical Society of the Pacific, 121,* 1296.

Najita, K. and Orrall, F.Q. (1970). White light events as photospheric flares. *Solar. Phys., 15,* 176-194.

Neidig, D. F. (1989), The importance of solar white-light flares. *Solar Phys., 121,* 261-269.

Neidig, D. F. and Cliver, E. W. (1983). A catalog of solar white-light flares, including their statistical properties and associated emissions, 1859 - 1982. *AFGL Technical Report.* U.S. Air Force Geophysics Laboratory, AFGL-TR- 83-0257 Report.

Ohki, K., and Hudson, H.S. (1975), Solar flare infrared continuum, *Solar Phys., 43,* 405-414.

Sakai, J. I., Nagasugi, Y., Saito, S. and Kaufmann, P. (2006). Simulating the emission of electromagnetic waves in the terahertz range by relativistic electron beams. *Astron. Astrophys., 457,* 313-318.

Shimabukuro, F.I. (1972). On the Temperature and Emission Measure of Thermal Radio Bursts. *Solar Phys., 23,* 169-177.

Shklovsky, J. (1964). The Inverse Compton Effect as a Possible Cause of the X-ray Radiation of Solar Flares. *Nature, 202,* 275-276.

Silva, A. V., Share, G. H., Murphy, R. J., Costa, J. E., de Castro, C. G., Raulin, J.-P. and Kaufmann, P. (2007). Evidence that Synchrotron Emission from Nonthermal Electrons Produces the Increasing Submillimeter Spectral Component in Solar Flares. *Solar Phys., 245,* 311-326.

Stein, W. A. and Ney, E. P. (1963). Continuum Electromagnetic Radiation from Solar Flares. *J. Geophys. Res., 68,* 65-81.

Trottet, G., Krucker, S., Lüthi, T. and Magun, A. (2008). Radio Submillimeter and gamma-Ray Observations of the 2003 October 28 Solar Flare. *Astrophys. J., 678,* 509-514.

Turon, P. J. and Léna, P. J. (1970). High resolution solar images at 10 microns: Sunspot details and photometry. *Solar Phys., 14,* 112-124.

White, S. M., Kundu, M. R., Bastian, T. S., Gary, D. E., Hurford, G. J., Kucera, T., Biejing, H. (1992). Multifrequency observations of a remarkable solar radio burst. *Astrophys. J., 384,* 656-664.

Xu, Y., Cao, W., Liu, C., Yang, G., Qiu, J., Jing, J., Denker, C., and Wang, H. (2004). Near-Infrared Observations at 1.56 Microns of the 2003 October 29 X10 White-light Flare. *Astrophys. J., 607,* L131-L134.

Xu, Y., Cao, W., Liu, C., Jing, J., Denker, C., Emslie, A.G. and Wang, H. (2006). High-resolution Observations of Multiwavelength Emissions During Two X-class White-Light Flares. *Astrophys. J., 641,* 1210-1216.

**Figure Captions**

**Figure 1** – Low-pass membrane filter transmission for wavelengths larger than about 3 µm, specially fabricated by Tydex LCC, for testing the 30 THz detector effective suppression of near-IR and visible radiation from the Sun. Upper panel shows the transmission for a wider wavelength range. Lower panel shows expanded scale near zero transmission at shorter near IR wavelengths.

**Figure 2** – Images of the GOES X class solar burst of October 27, 2014 (SOL2014-10-27T14:22), taken nearly the maximum of the time profiles at SDO EUV, HMI white-light, Hα and microwaves. In the top panels we have at left, the two white-light brightening, indicated by the arrows, at right their positions on the visible disk are indicated by the black arrows. Bottom panels show at left the Hα image, and the 30 THz brightening at right. All features coincide rather well in space. The bottom-left circle in the 30 THz image show the Airy disk, set by the photometric beam.

**Figure 3** – Time profiles for the October 27, 2014 solar flare (SOL2014-10-27T14:22), taken from RHESSI X-rays (top panels); SDO/HMI white-light images (the HMI "Ic" product) with cadence of one frame every 45 seconds at the top, compared to SST observations at 0.2 and 0.4 THz time profiles (for beams 2 and 5, respectively) at the bottom integrated over one second, presented in antenna temperature corrected for atmosphere transmission. One distinctive peak at about 14:22 UT is clearly identified at white light with corresponding impulsive bursts at 50-100 keV and 25-50 keV hard X-rays, 0.2 THz and at 0.4 THz.

**Figure 4** – The October 27, 2014 (SOL2014-10-27T14:22) emission spectrum from microwaves to THz frequencies at the maximum emission time exhibits the same spectral trends found for the two other 30 THz bursts analyzed before [*Kaufmann et al.,* 2013; *Miteva et al.,* 2015]. For this event the SST detections were well defined for the two sub-THz frequencies (with fluxes within uncertainties of about ±25 %).

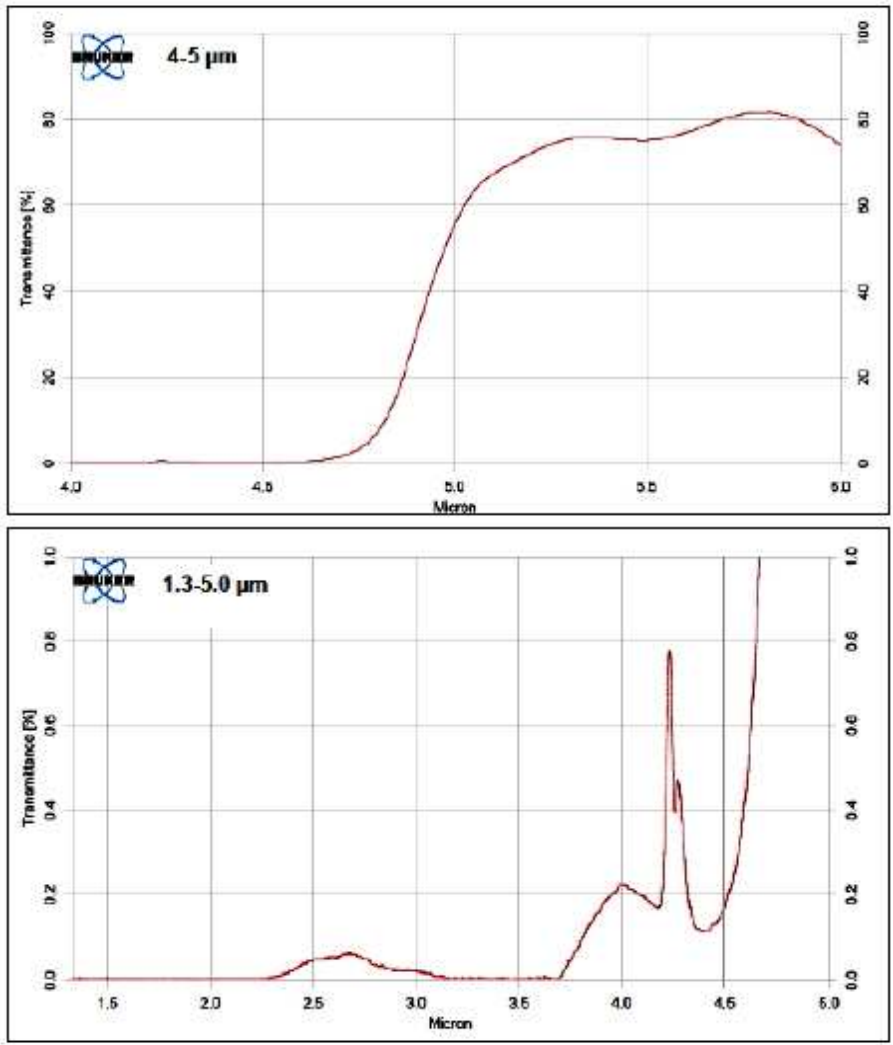

FIG 1

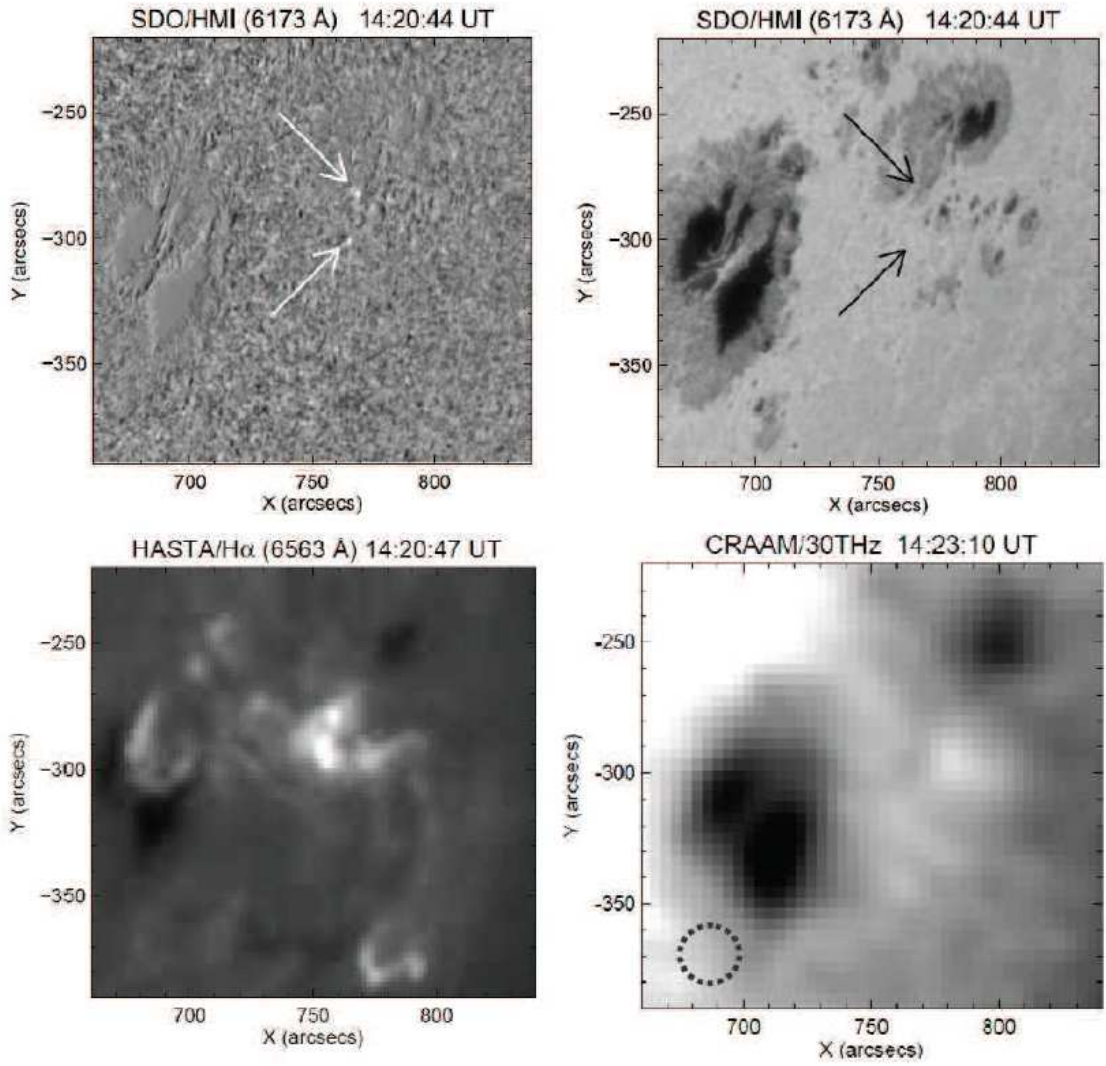

FIG 2

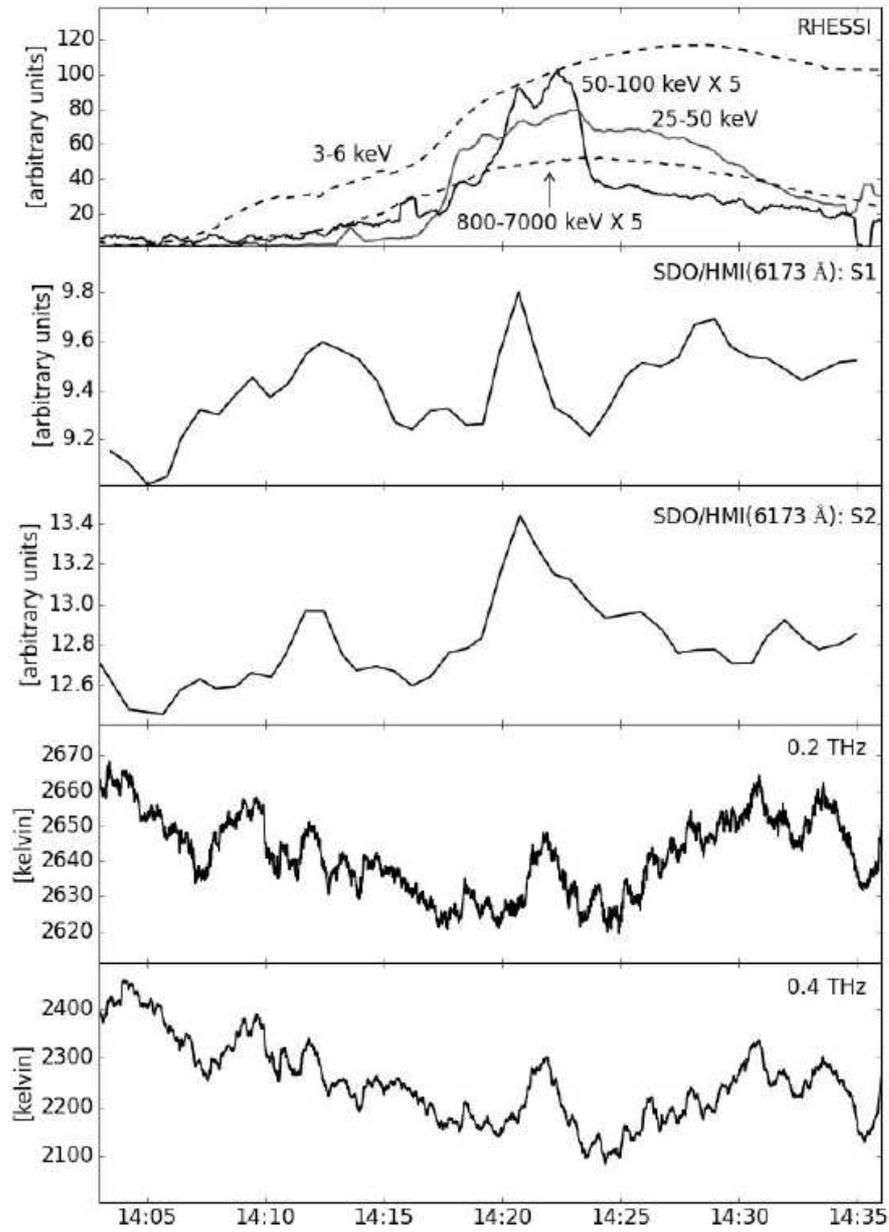

FIG 3

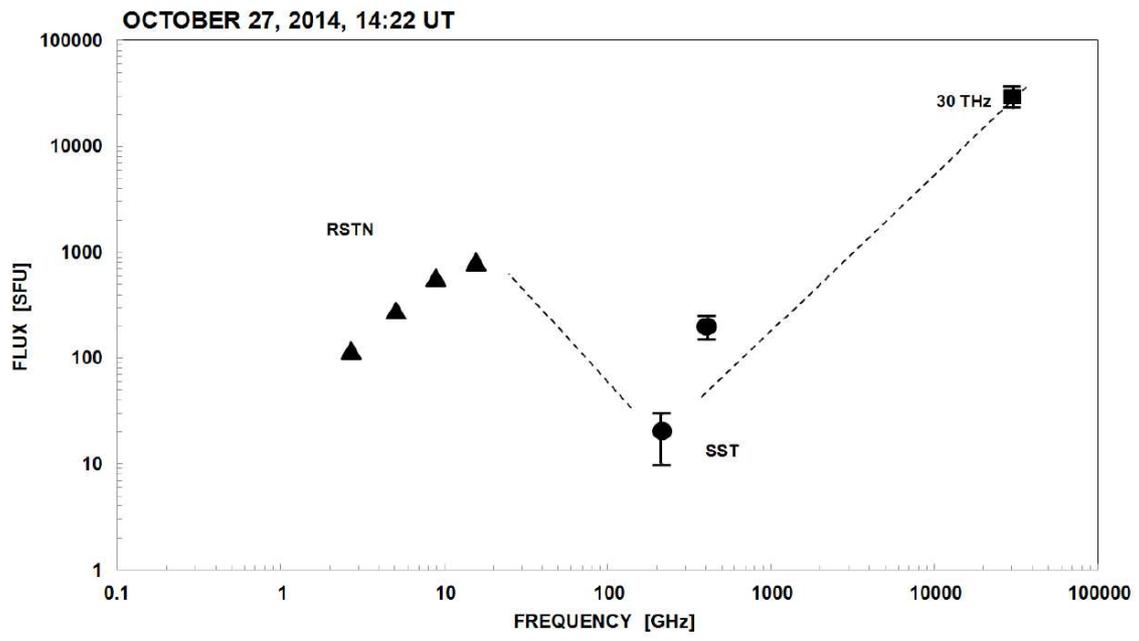

FIG 4